\def\includeSI{1}

\documentclass[prl,graphicx,reprint,floatfix,superscriptaddress]{revtex4-2}

\usepackage{float}
\usepackage{graphicx}% Include figure files
\usepackage{dcolumn}% Align table columns on decimal point
\usepackage{chemformula}
\usepackage{gensymb} % for e.g. \degree
\usepackage{amssymb} % for e.g. \square
\usepackage{hyperref}
\usepackage{upgreek}
\usepackage{siunitx}
\usepackage{eqtcommands}

\ifx\includeSI\undefined
    \newcommand{\suppinfo}{\cite{suppinfo}} 
    \newcommand{\suppSecNonlinear}{S7} 
\else
    \newcommand{\suppinfo}{(see Supp. Mat.)} 
    \newcommand{\suppSecNonlinear}{\ref{supp:nonlinear}} 
\fi

\begin{document}

%\preprint{}

\title{Spatial mapping of intrinsic and readout nonlinearities in a strongly-driven micromechanical membrane}



\author{Timo Sommer}
\affiliation{Department of Physics, TUM School of Natural Sciences, Technical University of Munich, Garching, Germany}
\affiliation{Munich Center for Quantum Science and Technology (MCQST), Munich, Germany}

\author{Agnes Zinth}
\affiliation{Department of Physics, TUM School of Natural Sciences, Technical University of Munich, Garching, Germany}
\affiliation{Munich Center for Quantum Science and Technology (MCQST), Munich, Germany}

\author{Aditya}
\affiliation{Department of Physics, TUM School of Natural Sciences, Technical University of Munich, Garching, Germany}

\author{Menno Poot}
\email{menno.poot@tum.de}
\affiliation{Department of Physics, TUM School of Natural Sciences, Technical University of Munich, Garching, Germany}
\affiliation{Munich Center for Quantum Science and Technology (MCQST), Munich, Germany}
\affiliation{Institute for Advanced Study, Technical University of Munich, Garching, Germany}

\date{\today}

\begin{abstract}
Recently, it was shown that strongly driven micromechanical resonators show mode shapes that strongly differ from the eigenmodes \cite{yang_PRL_spatial_modulation}. This raises the question of the origin of this nonlinear behavior. We measure the spatial dependence of the nonlinearities of high-stress micromechanical membranes. The mechanical nonlinearity is determined from the frequency response and is found to be independent of the probing location. It is the phase of the response that is instrumental in extracting that intrinsic nonlinearity. Our interferometric readout results in high-harmonics generation. These harmonics have a clear spatial profile that shows ring-like patterns resembling previous reports. These patterns are reproduced by a model of the displacement-dependent reflection signal in combination with motion amplitudes of the same order as the probing wavelength.
\end{abstract}

\pacs{}% insert suggested PACS numbers in braces on next line

\maketitle %\maketitle must follow title, authors, abstract and \pacs
In recent years, stressed micromechanical membranes have become an important platform for a wide variety of optomechanical experiments \cite{Zwickl_APL_membrane, adiga_APL_SiN_drum_Q_mode}. This ranges from applications in ultra-sensitive scanning probe microscopy \cite{Haegl_PRAppl_force_sensing, Eichler_MQT_review_Q} and detection of radio waves \cite{bagci_nature_radio_wave_detection}, determining material properties \cite{sommer_APL_membrane_AlN}, to study cavity optomechanical backaction \cite{thompson_nature_cavity_membrane} including radiation pressure shot noise \cite{purdy_science_RPSN}, topological energy transfer \cite{xu_nature_exeptional_points} and nonreciprocal dynamics \cite{xu_nature_nonreciprocal}, all the way to radiative heat transfer mediated by Casimir fluctuations \cite{fong_nature_heat_Casimir}.
Although these experiments mostly utilized the out-of-plane vibrations of the membranes as simple harmonic oscillators, also the spatial profile of these vibrations is important, for example in the observation of hybridization of degenerate eigenmodes \cite{sommer_ST_xtalk} or the analysis of clamping losses in the phonon-tunneling framework \cite{Wilson-Rae_PRL_highQ_membranes, Chakram_PRL_membrane_image}. In this context, mode mapping is a powerful tool to visualize - and even to quantitatively analyze - the spatial structure of eigenmodes. It is also instrumental for understanding the linear dynamics of complex nano- and micromechanical structures \cite{garcia_NL_imaging_graphene, etaki_natphys_squid}. 
%Mode visualization is typically done in the linear regime, where the locally-probed displacement is proportional to the applied force and the results are readily interpreted. For example, it was found that even very complex-looking membrane modes can be decomposed into a linear combination of nearly-degenerate eigenmodes \cite{hoch_MM_mode_mapping}.
The nonlinear regime is far less explored. Recently, % persistent responses \cite{yang_PRL_persistent_response} and 
unexpected emerging spatial structures were reported in strongly-driven micromechanical membranes \cite{yang_PRL_spatial_modulation}, but the origin of these signals remains puzzling. % This puzzle is solved in this work with a careful study of the optical readout mechanism. , which is resolved in this work. 
Here we show, that spatially resolving the different nonlinearities is essential to solving that puzzle and other intriguing observations. %, it is essential to spatially resolve the different nonlinearities.
For this, we extend our efficient optomechanical mode mapping technique \cite{hoch_MM_mode_mapping, sommer_ST_xtalk} to spatially map nonlinearities in high-stress silicon nitride (SiN) membranes. We locally probe nonlinear responses of both a high and low quality factor ($Q$) mode in a single membrane. % , the intrinsic and readout nonlinearities are distinguished. 
The former displays a regular Duffing response, but the latter mode has strongly distorted resonances. We show that in this case, the \emph{phase} can be used to determine the Duffing parameter $\alpha_y$. Even though the driven response, and with it $\alpha_y$, varies strongly over the membrane, the extracted mechanical nonlinearity of both modes is found to be position independent. 
For the low-$Q$ mode, %requires a much larger amplitude to shift the resonance frequency by a linewidth \cite{Roy_science_sensor_damping} and
nonlinear readout dominates and overtones of the driving frequency with a distinct spatial pattern appears. Maps of the first six harmonics show characteristic annuli of alternating positive and negative regions.
These observations are captured by a displacement-dependent reflectivity model. The dynamics are also studied in the time domain, where complex waveforms are observed. % The combined results provide insights into the interplay between intrinsic and readout nonlinearities in nano- and micromechanical devices. 

\begin{figure}[!hbt]
  \includegraphics[width=1.0\columnwidth]{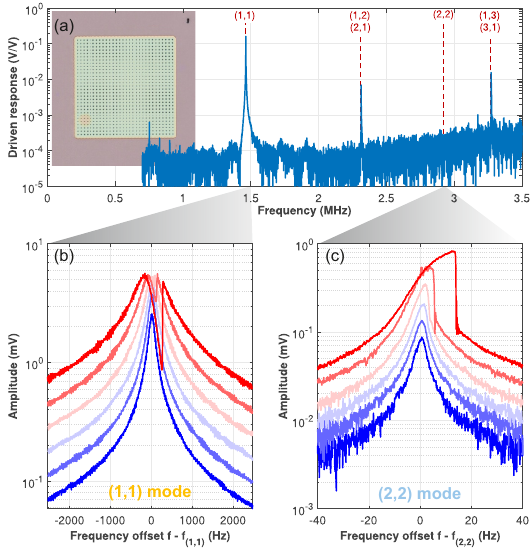}
  \caption{
  (a) Overview driven response of a $270 \times 270 \un{\mu m}$ SiN membrane with the modes indicated. The inset shows an optical micrograph of the membrane. 
  (b) Driven response of the (1,1) mode for varying driving powers from -30 to -10 dBm (blue to red). The eigenfrequency of the fundamental mode is $f_{(1,1)} = 1.4660 \un{MHz}$. Fits in the linear regime yield $Q_{(1,1)} = 11\mathrm{k}$, corresponding to a linewidth of $135 \un{Hz}$.
  (c) Driven response of the (2,2) mode for varying driving powers from -34 to -14 dBm. The (2,2) mode is around $f_{(2,2)} = 2.9323 \un{MHz}$ and has $Q_{(2,2)} = 742\mathrm{k}$, yielding a linewidth of $3.95 \un{Hz}$. At higher driving powers, a stiffening Duffing nonlinearity is visible.
\label{fig:intro}}
\end{figure}

Highly stressed membranes are a well-understood model system and we fabricate these by defining holes in SiN and removing the silicon oxide underneath \cite{adiga_APL_SiN_drum_Q_mode, Wilson-Rae_PRL_highQ_membranes, cha_natnano_phononic_electro, kim_NL_buckling_waveguide} 
(for details, see the Supp. Mat. at the end of this document).
This approach has the benefit that the interference between reflections from the Si substrate and the membrane is an easily understood method to measure the local membrane displacement \cite{hoch_MM_mode_mapping}. Many observations we report in this Letter were observed in multiple devices and chips, but we focus on a single membrane that was characterized in detail. A micrograph is shown as inset in Fig.~\ref{fig:intro}(a), which shows the driven response of the membrane. In this overview, several sharp resonances can be observed - the eigenmodes of the membrane. The frequency ratios closely follow those of an ideal membrane (dashed lines) so that the modes can be identified. %This is confirmed by mode maps \suppinfo.
The modes are labeled $i=(m,n)$ where $m$ and $n$ count the number of antinodes in the $x$ and $y$ direction \cite{yu_PRL_membrane_Al}. In the following, we focus on the (1,1) and the (2,2) modes, which do not show any hybridization \suppinfo. Zooms at higher resolution allow the extraction of the resonance frequencies $f_i$ and the quality factor $Q_i$ of each mode. 
% The exact value of $Q$ depends on 
%the particular membrane and mode under study. These variations are %explained in the literature due to different the amount of 
% coupling to the substrate modes \cite{Wilson-Rae_PRB_phonon_tunneling, Wilson-Rae_PRL_highQ_membranes, Joeckel_PRL_spectroscopy_Q_SiN_membrane, Chakram_PRL_membrane_image}. 
%It is thus possible to have modes in a single membrane with widely different quality factors. 
%Note that these modes can be unambiguously identified just based on their frequency and that these are not prone to mode hybridization as is often observed for degenerate modes, such as the (1,2)/(2,1) and the (1,2)/(2,1) doublets \cite{sommer_ST_xtalk, Chakram_PRL_membrane_image}. For this particular membrane, 
$Q_{(1,1)} \sim 1\times 10^4$ is rather low, indicating a strong coupling to lossy substrate modes \cite{Wilson-Rae_PRB_phonon_tunneling, Wilson-Rae_PRL_highQ_membranes, Joeckel_PRL_spectroscopy_Q_SiN_membrane, Chakram_PRL_membrane_image}, whereas $Q_{(2,2)} \sim 7\times 10^5$ is among the highest values observed in our membranes. The line widths $w_i = f_i/Q_i$ are thus very different [the (2,2) resonance is so narrow that it is not visible in Fig.~\ref{fig:intro}(a)], implying that the critical amplitude, i.e. the motion amplitude where the Duffing response becomes vertical \cite{Roy_science_sensor_damping} \suppinfo, is also very different for these two modes. As will be shown below, this causes the (2,2) mode to be dominated by intrinsic mechanical nonlinearities, whereas the (1,1) resonances are strongly affected by the optomechanical readout. To study this, the modes are excited with varying strengths;
Figure~\ref{fig:intro}(b) and (c) show how their responses evolve. At low power (blue), both modes show the symmetric response of a harmonic oscillator. For stronger driving, however, the shape changes. The (2,2) mode in (c) shows a steepening of the response all the way to vertical jumps. This ``shark fin'' is typical for driven Duffing resonators where besides a linear term $-ku$ also cubic (and possibly quadratic \cite{postma_APL_dynamicrange_NW, lifshitz_coupled_NL}) terms in $u$ contribute to the restoring force \suppinfo. In contrast, the (1,1) mode shows the emergence of a dip in the response, and the maximum signal seems to be clamped at $\sim 5 \un{mV}$. This is not expected for a Duffing resonator, but spatial mapping can provide insight into the origin of these observations.
\begin{figure}[tb]
  \includegraphics[width=1.0\columnwidth]{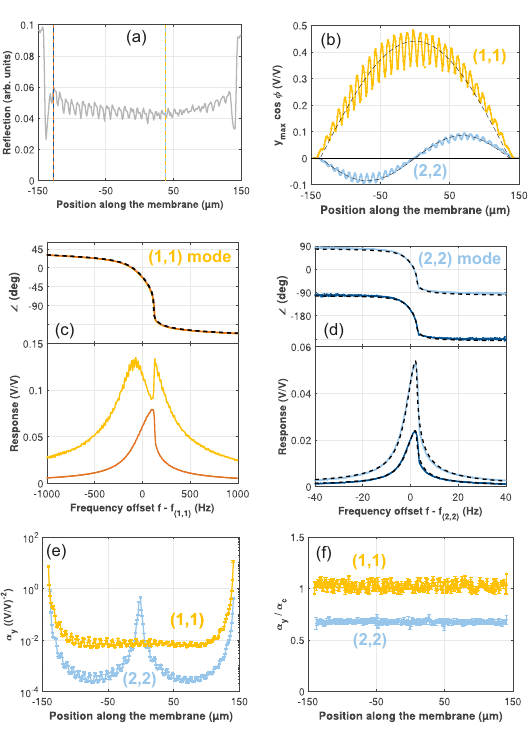}
  \caption{
  (a) Measured reflectivity across the membrane. The supported region has higher reflectivity than the membrane. % at the used wavelength of 633 nm. The holes show up as fluctuations. 
  The dashed lines indicate the locations where the responses in (c) and (d) %and the power dependence in (f) 
  were measured.
  (b) Measured mode shapes.
  % for the (1,1) and the (2,2). 
  The dashed lines indicate the theoretical mode shapes.
  (c) Magnitude $|y(f)|$ (bottom) and phase $\angle y(f)$ (top) of the driven response of the (1,1) mode at two different locations [see (a)]. The dashed line indicates the fits.
  (d) same as (c), but for the (2,2) mode.
  (e) Extracted $\alpha_y$.
  (f) Normalized Duffing parameter $\alpha_y/\alpha_c$ across the membrane. 
  % The inset shows the dependence of the normalized Duffing parameter on the driving power at the positions indicated in (b). The black line shows the expected quadratic dependence.
  In all panels, yellow and blue shades indicate the (1,1) and (2,2) mode, respectively. Darker (lighter) colors are taken near the edge (center) as indicated in (a). Error bars indicate the fit uncertainty. 
  % NWA power -14 dBm for (1,1), -22 for (2,2). HP NWA
  \label{fig:linecuts}}
\end{figure}

%Figure~\ref{fig:intro} was measured at one specific location. 
Figure~\ref{fig:linecuts} shows how the responses change while scanning over the membrane. For reference, Fig.~\ref{fig:linecuts}(a) shows the reflectivity $R$. The suspended membrane has a lower $R$ compared to the edges. The oscillations in $R$ arise due to the etch holes, which are also visible in the measured mode shapes shown in Fig.~\ref{fig:linecuts}(b). Their overall profile matches the theoretical shape of the (1,1) and (2,2) modes. 
% (2d maps can be seen in \suppinfo)
To drive the modes nonlinear, the excitation is increased. Figure~\ref{fig:linecuts}(c),(d) show the resulting frequency response $y(f)$ at two locations: near the edge and close to the center. At both locations, the (2,2) response is described well by the Duffing model and the fits yield the Duffing parameter $\alpha_y$ \suppinfo. This holds for both the magnitude (bottom) and the phase (top panel). The offset of $180\degree$ in its phase is due to the opposite motion at the two locations. 
In contrast, the magnitude of the (1,1) mode shows a Duffing response near the edge, but again a strongly distorted response with a dip near the center [cf. Fig.~\ref{fig:intro}(b)]. Still, the \emph{phase} looks regular and matches the phase of the Duffing response. This important result, which allows extraction of $\alpha_y$ for the distorted resonances, is formalized in Sec.~\suppSecNonlinear. Fitting is done for all responses measured along the membrane. The resulting $\alpha_y$ in Fig.~\ref{fig:linecuts}(e) % \footnote{For the (1,1) the ratio $\alpha_y/\alpha_c$ is obtained from the fitted phase response and converted back to $\alpha_y$ using the maximum amplitude in the linear regime \suppinfo}. 
varies strongly across the membrane; for both modes $\alpha_y$ is largest near the edges and for the (2,2) mode it also shows a strong increase near the nodal line. However, it turns out that this rise in $\alpha_y$ is due to the smallness of the response there. When normalizing the Duffing parameter to the critical value $\alpha_c$, i.e. the value of $\alpha_y$ where the response just becomes vertical, the profiles in Fig.~\ref{fig:linecuts}(f) are obtained. Despite the two orders of magnitude change in the underlying $\alpha_y$, the normalized Duffing parameter $\alpha_y/\alpha_c$, quantifying the intrinsic mechanical nonlinearity of the modes \suppinfo, is independent of the position. This demonstrates that the mechanical nonlinearities of our membranes are indeed a property of their eigenmodes and are, thus, independent of the probing position. 
\begin{figure*}[tb]
  \includegraphics[width=1.0\textwidth]{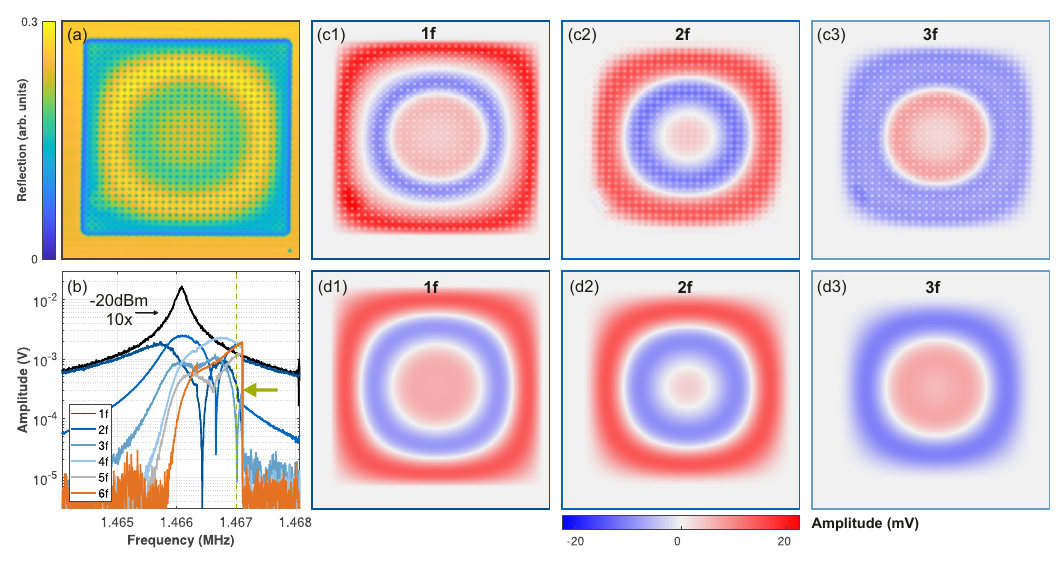}
  \caption{
  Maps and spectra of the harmonics at 0 dBm driving power.
  (a) Reflectivity map of the membrane. 
  (b) Driven response of the (1,1) mode and the co-recorded harmonics. The green dashed line indicates the driving frequency used for the measurements in panels (a) and (c); the green arrow indicates the jump from the high to the low branch. For comparison also the scaled amplitude (1f) for lower power is shown (black curve). 
  (c1)-(c3) Amplitude maps of the 1st to 3rd harmonic of the driving frequency. (d1)-(d3) Harmonic maps calculated with our readout model \suppinfo. The size of all maps is $350 \times 350 \un{\mu m^2}$. 
  Note that the data in (a) and (c) were recorded during the same measurement. The data in (b) is an earlier measurement, and the measurement setup was optimized in between so that the amplitudes cannot be compared directly.
  \label{fig:maps}}
\end{figure*}
Still, the question remains: what causes the distorted resonances in Fig.~\ref{fig:intro}(b) and, importantly, what spatial structure does this have? As we will show, this is caused by nonlinearities in the readout. 
Sending a sinusoidal signal (in our case, the membrane motion) through a nonlinear element (the optomechanical detection) generates harmonics of the fundamental frequency. % The two toy models in the Supp. Mat. \suppinfo show that this strongly depends on the signal amplitude. 
It is thus important to not only record the signal at the driving frequency $f$, but also at its harmonics ($nf$). %This can be done for six harmonics simultaneously with the lock-in amplifier (Zurich Instruments HF2). 
We select the fundamental frequency ($n=1$) and the first 5 overtones ($n=2..6$) and drive the (1,1) mode strongly at 0 dBm. The signal amplitude can be changed via the probing position or via the driving frequency. 
For the former, it was already shown that the motion was small near the edges (dark curves in Fig.~\ref{fig:linecuts}) and near nodal lines, and maximal at the anti-nodes of the mode (light curves). 
For the latter, a large detuning means a small amplitude whereas driving on resonance results in a high amplitude as shown in Fig.~\ref{fig:maps}(b). Far detuned, the first harmonic (dark blue) follows the appropriately scaled linear response (black), and the overtones are small; only $n=2$ is visible above the noise floor. When approaching the resonance frequency, the harmonics start to appear and these rise with different slopes. At the same time, the fundamental tone decreases below that of the linear response. At specific frequencies, sharp dips appear in the harmonics, where that signal vanishes. Upon further increasing the frequency, the response of all harmonics stays high, followed by a sharp drop [green arrow in Fig.~\ref{fig:maps}(b)] where the Duffing resonator jumps to the low-amplitude state. After this, the fundamental harmonic follows the linear response again, and the overtones are small. The observed frequency dependence of the harmonics is explained well by considering the model for the nonlinear readout of the motion as detailed in Sec.~\suppSecNonlinear.

Next, maps are acquired using our mode mapping technique \cite{hoch_MM_mode_mapping}, but this time for the different harmonics instead of for different modes as shown in Fig.~\ref{fig:maps}(c). 
Panel (c1) shows the first harmonic and can thus be compared with the regular mode maps (see Fig.~S3 in the Supp. Mat.). With strong driving, the map looks different from the (1,1) mode: instead of a single maximum in the center, now a ring-shaped pattern appears. After the initial increase in amplitude when going away from the clamping sides (darker red), there is a reduction in the signal (lighter red). This corresponds to the dip that appeared in the frequency response. When moving further towards the center, an almost circular line of zero amplitude can be observed, followed by a blue ring. There, the signal is in anti-phase compared to the motion \cite{hoch_MM_mode_mapping}. Even closer to the center, the signal is again in phase but with a strongly reduced amplitude. Similar patterns with annular regions of in- and anti-phase motions are also observed in the higher harmonics of the drive and also appear in the reflectivity map [Fig.~\ref{fig:maps}(a)], which contains the $n=0$ harmonic. 
% , which can be viewed as the $n=0$ harmonic. 

\begin{figure}[tb]
  \includegraphics[width=1.0\columnwidth]{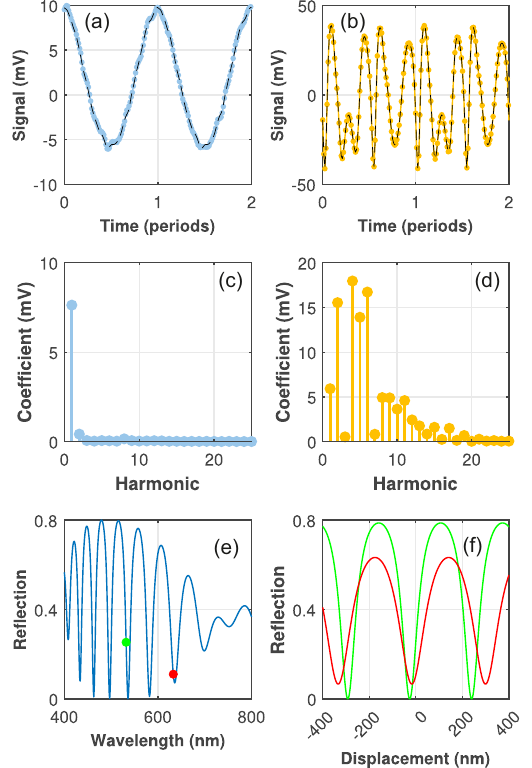}
  \caption{
  (a), (b) Time traces of the driven motion for low (a) and high amplitude (b). The two periods of the drive correspond to 1.37 and 1.36 $\un{\mu s}$, respectively.  
  (c), (d) Decomposition of the signals from (a), (b) into harmonics. The reconstructed signal is shown as the dashed lines in (a), (b).  
  (e) Calculated reflectivity spectrum $R(\lambda)$ for an air/SiN/air/Si multilayer \suppinfo. 
  (f) Calculated $R(u)$ for $\lambda = 632.8 \un{nm}$ (red) and 532 nm (green), as indicated in (e). 
  \label{fig:time}}
\end{figure}

The different harmonics all have similar maximum magnitudes of about 20 mV, which is large enough to study the signal in the time domain. Figure~\ref{fig:time}(a) and (b) show measured time traces for small and large amplitudes, respectively. For the small amplitude, the detected signal is close to a sinusoidal oscillation and almost all power is carried by the $n=1$ component [Fig.~\ref{fig:time}(c)]. The readout is thus close to linear. This is in stark contrast with the high amplitude case, where a complex waveform with many minima and maxima per oscillation period is observed. % The Fourier decomposition of the signal into the harmonics in 
Figure~\ref{fig:time}(d) shows that, in analogy with mechanical frequency combs \cite{Jong2023_mechOvertoneFreqComb}, it contains many harmonics, e.g. $n = 21$ still is $ > 1\%$ of the maximum, which occurs at $n=4$. In contrast to the combs in Ref.~\cite{Jong2023_mechOvertoneFreqComb}, in our case, the \emph{readout} is strongly nonlinear. Also, our combs are strongly position-dependent cf. \figref{fig:maps}. 
%The question remains what causes the nonlinearity in the readout? In our case, the motion changes the wavelength-dependent reflected laser light \cite{hoch_MM_mode_mapping, suppinfo}
Our interferometric detection~\cite{hoch_MM_mode_mapping} uses the displacement dependence of the reflectivity $R(u,\lambda)$. Figure~\ref{fig:time}(e) and (f) shows how $R$ changes with wavelength $\lambda$ and displacement $u$. Small changes in $u$ result in proportional changes in $R$. 
If the amplitude grows, $R(u)$ is no longer linear as described in e.g. Ref.~\cite{dolleman_APL_calibration_NL_readout}, ultimately resulting in sweeping over multiple minima and maxima during a single oscillation, exactly as observed in Fig.~\ref{fig:time}(b). By modeling $R(u)$ and applying this to analytical (1,1) mode shape \suppinfo gives the reflectivity maps of Fig.~\ref{fig:maps}(d1-3), as well as the calibrated amplitude \cite{dolleman_APL_calibration_NL_readout}, which is \SI{446}{nm}. These harmonic maps also show characteristic annuli and match their experimental counterparts (c1-3) well. In the Supp. Mat., we also show this for the first six harmonics and derive the nonlinear readout model in detail.
%and our models for the nonlinear readout \suppinfo capture not only the harmonics' frequency dependence but also their spatial profile and the dynamics in the time domain.

In conclusion, we have spatially mapped a strongly driven micromechanical membrane. We show that the mechanical nonlinearity is a mode property and does not depend on the measurement position. The observed ring-like spatial patterns are not an effect of the mechanical nonlinearities but originate from the readout. We show that motions exceeding the linear range of $R(u)$ result in complex waveforms and high harmonics; our model captures all the observed effects. These results enable a deeper comprehension of the complex nonlinear dynamics of a wide variety of micromechanical resonators and readout methods.

\begin{acknowledgments}
We thank Elke Scheer and Gianluca Rastelli for discussions, Samer Houri for feedback on the manuscript, and David Hoch and Nirav Mange for assistance with nanofabrication. Funded by the German Research Foundation (DFG) under Germany's Excellence Strategy - EXC-2111-390814868 and TUM-IAS, funded by the German Excellence Initiative and the European Union Seventh Framework Programme under grant agreement 291763.
\end{acknowledgments}

\bibliography{nonlinear.bbl}

\newcommand{\figintro}{\ref{fig:intro}}
\newcommand{\figlinecuts}{\ref{fig:linecuts}}
\newcommand{\figmaps}{\ref{fig:maps}} 
\newcommand{\figtime}{\ref{fig:time}}

\ifx\includeSI\undefined
\else
    \def\calledSI{1}
    %\appendix
\ifx\calledSI\undefined
   \documentclass[prl,graphicx,reprint,floatfix,superscriptaddress]{revtex4-2}

    \usepackage{float}
    \usepackage{graphicx}% Include figure files
    \usepackage{dcolumn}% Align table columns on decimal point
    \usepackage{chemformula}
    \usepackage{amssymb} % for e.g. \square
    \usepackage{hyperref}
    \usepackage{upgreek}
    \usepackage{siunitx}
    \usepackage{eqtcommands}

    \newcommand{\figintro}{1}
    \newcommand{\figlinecuts}{2}
    \newcommand{\figmaps}{3}
    \newcommand{\figtime}{4}

    \draft % marks overfull lines with a black rule on the right

    \begin{document}
\fi

\setcounter{secnumdepth}{2} % https://tex.stackexchange.com/questions/488302/referencing-an-appendix-while-using-prl-style
\setcounter{section}{0}
\renewcommand\thesection{S\arabic{section}}
\setcounter{figure}{0}
\renewcommand\thefigure{S\arabic{figure}}
\setcounter{equation}{0}
\renewcommand\theequation{S\arabic{equation}}
\setcounter{table}{0}
\renewcommand{\thetable}{A\arabic{table}}

\newpage
\begin{huge}
\noindent Supplementary~Material
\end{huge}

\tableofcontents

\section{Fabrication}\label{supp:fabrication}
Samples were fabricated from commercially acquired wafers with a \SI{315}{nm} thick stoichiometric LPCVD SiN layer on top of a $3.17\un{\mu m}$ thick silicon oxide (SiOx) layer, which is again on top of the silicon carrier. The SiN has a tensile stress of about \SI{1100}{MPa}. The wafer was diced into $6 \times 10 \un{mm}$ dies. After dicing, a chip was cleaned and spin-coated with ZEP520A resist. 120 membranes with different sizes are then made by defining release holes using electron-beam lithography followed by reactive ion etching and resist stripping. The holes expose the underlying oxide and by immersing the chip into buffered hydrofluoric acid (BHF), the exposed SiOx is etched isotropically, resulting in circularly expanding ``drums'' originating at the etch holes. The release etching is done for about \SI{130}{min.}, resulting in about $10\un{\mu m}$ radial etching of the SiOx. The release holes are arranged to ensure complete clearance of the SiOx in the area between the holes, resulting in the formation of a fully suspended SiN membrane \cite{adiga_APL_SiN_drum_Q_mode,hoch_MM_mode_mapping}. Note that the finite selectivity of BHF between SiN and SiOx results in slightly tapered profiles of the SiN \cite{sommer_APL_membrane_AlN}. After the release, the chip is dried using a critical point dryer. Finally, the chip is glued onto a piezo-electric element for actuation and mounted in a vacuum chamber as shown in Fig.~\ref{fig:setup}. All measurements are performed in vacuum ($\lesssim \SI{e-4}{mbar}$) so that air damping can be neglected.

All membranes on the chip were characterized in the context of determining the mechanical properties of aluminum nitride film \cite{sommer_APL_membrane_AlN}. This chip, however, does not have any AlN on it. The number of release holes and their distance determine the geometry of the membrane. The device in the main text (named ``J10'') has 30 by 30 holes with their centers spaced by $8.56\un{\mu m}$ resulting in a membrane size of $270\un{\mu m}$. An optical micrograph of the membrane was shown in Fig. 1 of the main text. 

\section{Measurement setup}\label{supp:setup}
\begin{figure}[tb]
  \includegraphics[width=1.0\columnwidth]{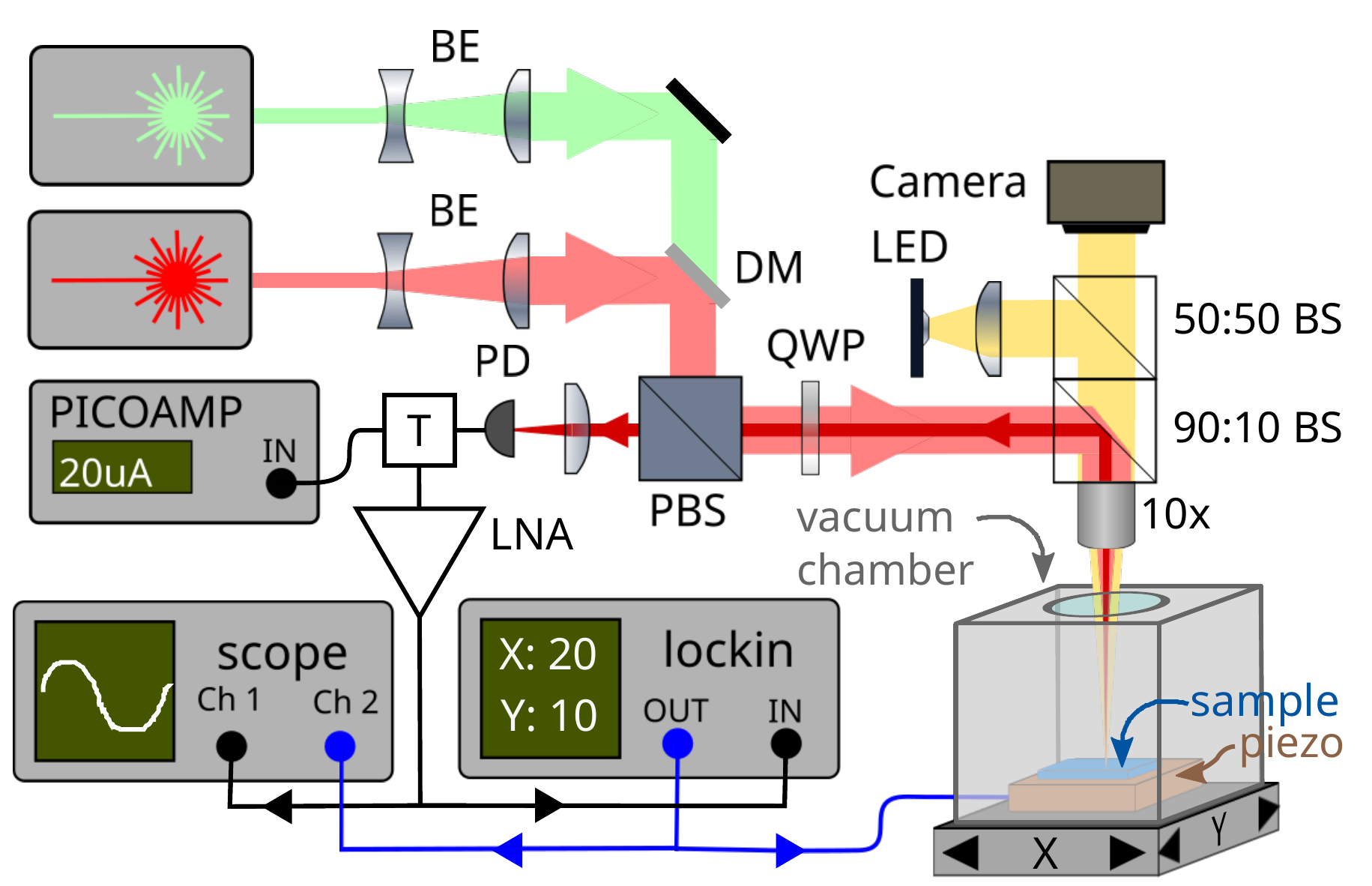}
  \caption{
  Schematic of the measurement setup. Two different lasers can be used for measuring the driven response of the membranes that are mounted on a piezo actuator and placed in the vacuum chamber. For clarity, the green laser light path is not shown in full. BE: beam expander, DM: dichroic mirror, (P)BS: (polarizing) beam splitter, QWP: quarter wave plate, PD: photodetector, LNA: low-noise amplifier, T: bias tee, LED: light emitting diode for illumination.
  Adapted from Ref.~\cite{hoch_MM_mode_mapping}.
  \label{fig:setup}}
\end{figure}

\begin{figure}[tb]
  \includegraphics[width=1.0\columnwidth]{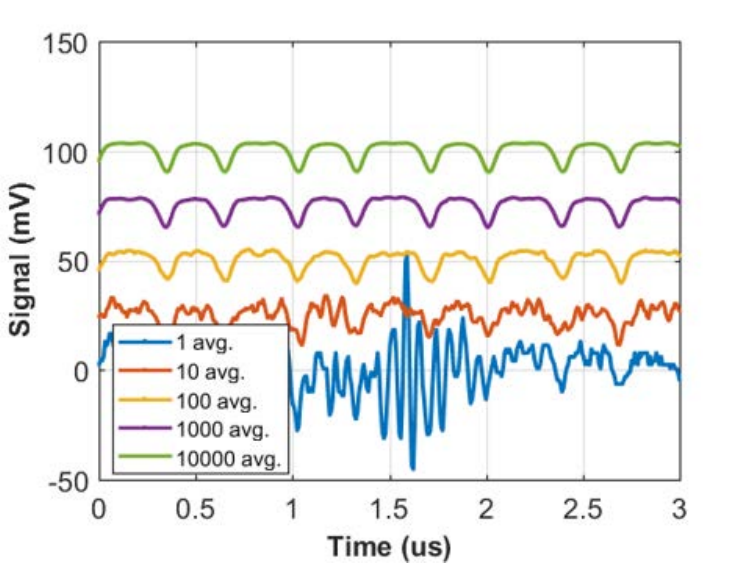}
  \caption{
  Time traces with different numbers of averages (offset for clarity). Without averaging (blue), the mechanical signal is buried in interference, but with more averaging, it starts to appear. The curves with 1000 and 10000 averages are virtually indistinguishable. Note that the signal amplitude remains the same, indicating that the triggering on the drive signal works well.
  \label{fig:avg}}
\end{figure}

The setup is shown schematically in Fig.~\ref{fig:setup}. Compared to our previous work in Refs.~\cite{hoch_MM_mode_mapping, sommer_ST_xtalk} the setup is extended with a green laser (532 nm, Onefive Katana 05/06). However, unless mentioned otherwise, a red HeNe laser (633 nm, Melles Griot 05-LHP-141) was used. The measurements were mostly done with a lock-in amplifier (Zurich Instruments HF2). A network analyzer (HP 4396A, not shown) was only used for the measurements in Fig. 1; the rest of the driven responses were measured using the lock-in amplifier. For the time traces, an oscilloscope (PicoScope 5442D) was used. The membrane motion was captured on Channel 1 (1024 samples, 125 MSa/s, 200 mV range) and averaged 1000 times. Triggering was done on the driving signal (blue) generated by the lock-in amplifier (connected to channel 2 of the oscilloscope). Figure~\ref{fig:avg} illustrates the averaging.

During the measurements, the temperature of the sample stage in the vacuum chamber was stabilized using a digital proportional-integral (PI) controller, regulating the amount of heating of a Peltier element via a programmable power supply.


\section{Mode maps of the first six modes}\label{supp:modemaps}
\begin{figure*}[tb]
  \includegraphics[width=1.0\textwidth]{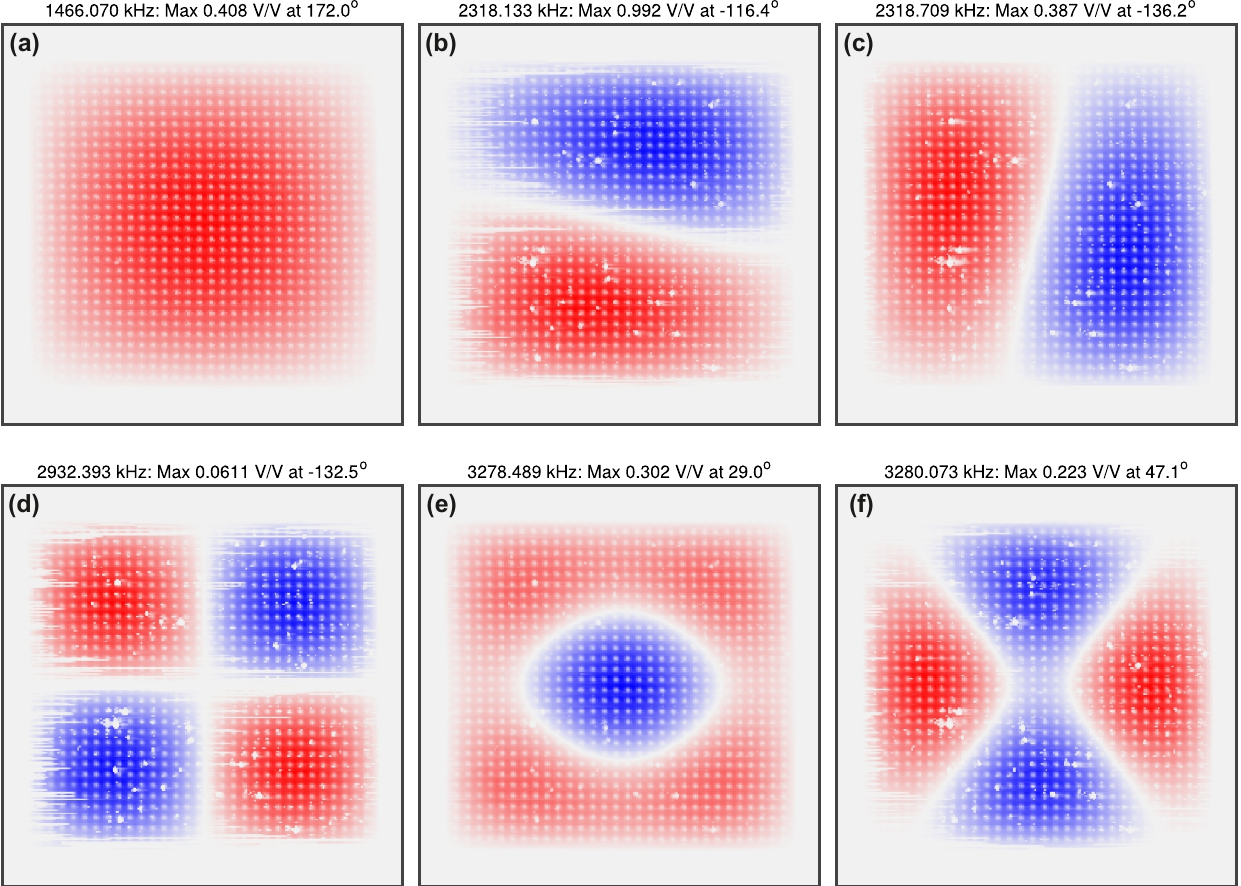}%Figures/nonlinear_supp_map_6modes.pdf
  \caption{
  Measured mode maps of the first six modes of the membrane \cite{hoch_MM_mode_mapping, sommer_ST_xtalk}. The frequency and maximum transmission are indicated above the panels, as well as the analyzer angle. The driving powers of the 6 modes were -32, -42, -35, -21, -30, and -32 dBm, respectively. % WSN06_55\J10_055_PLLmm_xsteps_ysteps.zi
  This corresponds to a weakly nonlinear drive ($\alpha_y/\alpha_c \sim 0.4$) for modes 2 to 6 and a linear drive for mode 1 (1,1). The color scale is different between the panels and the maps show an area of $325 \times 325 \un{\mu m^2}$.
  \label{fig:mode_maps}}
\end{figure*}

Mode maps of the first six modes of the membrane are shown in Fig.~\ref{fig:mode_maps}.  
The (1,1) and (2,2) modes are clearly identifiable. The (1,2) and (2,1) modes are also recognizable, although they have a non-straight nodal line, indicating weak mode mixing~\cite{sommer_ST_xtalk}. The last two modes are close to the even and odd superpositions of the (1,3) and (3,1) modes. White spots and stripes in the maps are regions where the PLL was unstable. %This work focuses on the (1,1) and (2,2) modes and more detailed data on the mode maps of these two modes are shown in Fig.~\ref{fig:map11} and \ref{fig:map22} respectively (all three mode-map figures are from the same measurement).

%
%\begin{figure*}[tb]
%  \includegraphics[width=1.0\textwidth]{Figures/nonlinear_supp_map_11.pdf}
%  \caption{
%  Mode maps measurement of the (1,1) mode at -32 dBm. The contours indicate lines of constant reflectivity (``detector 1''). The map $S(X,Y)$ is complex-valued with a magnitude $|S|$, real part containing the actual mode map, imaginary part containing the error signal, and the phase $\angle S$. The last panel shows the frequency shift by the PLL during acquisition. The analyzer angle is indicated. See Refs.~\cite{hoch_MM_mode_mapping, sommer_ST_xtalk} for more details.
%  \label{fig:map11}}
%\end{figure*}
%
%\begin{figure*}[tb]
%  \includegraphics[width=1.0\textwidth]{Figures/nonlinear_supp_map_22.pdf}
%  \caption{
%  Mode maps measurement of the (2,2) mode at -21 dBm. The contours indicate lines of constant reflectivity (``detector 1''). The map $S(X,Y)$ is complex-valued with a magnitude $|S|$, real part containing the actual mode map, imaginary part containing the error signal, and the phase $\angle S$. The last panel shows the frequency shift by the PLL during acquisition. The analyzer angle is indicated. See Refs.~\cite{hoch_MM_mode_mapping, sommer_ST_xtalk} for more details.
%  \label{fig:map22}}
%\end{figure*}

\section{Comparing red and green readout}\label{supp:redgreen}
\begin{figure}[tb]
  \includegraphics[width=1.0\columnwidth]{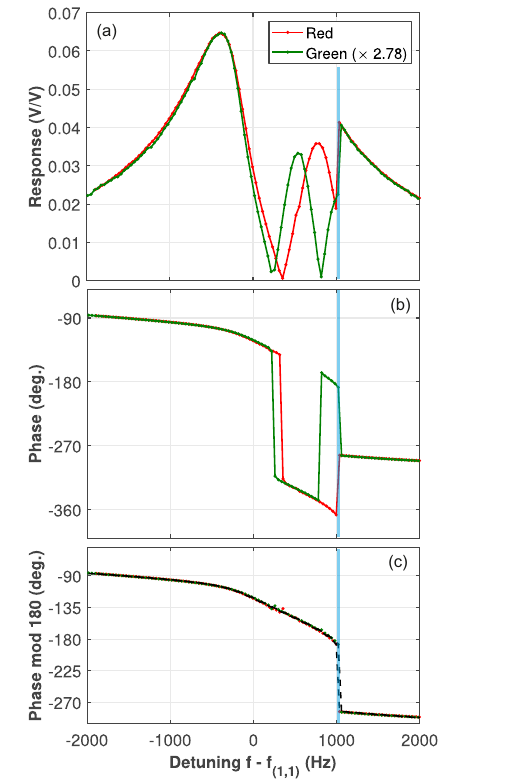}
  \caption{
  Magnitude (a) and phase (b) of the first harmonic measured with the red and the green laser while sweeping the driving frequency around $f_{(1,1)}$. The excitation power was 0 dBm. (c) Phase from (b) modulo $\pi$ together with the fitted phase response (dashed lines). The data in (c) is unwrapped. The location of the jump from the high to the low amplitude branch of the Duffing oscillator is indicated in blue.
  \label{fig:supp:red_green}}
\end{figure}

\begin{figure*}[tb]
  \includegraphics[width=1.5\columnwidth]{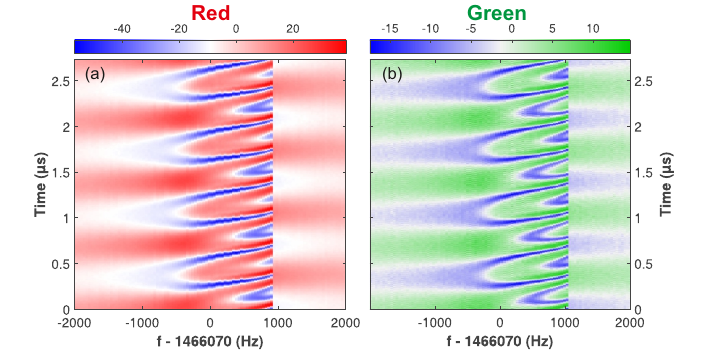}
  \caption{
  Measured time traces for approximately 4 oscillation periods as a function of the driving frequency for the red (a) and the green laser (b). The color scales are in mV. % nonlinear_compare_red_green.m
  \label{fig:supp:red_green_time}}
\end{figure*}

As mentioned in Sec.~\ref{supp:setup}, two different lasers can be used to detect the membrane's vibrations. The readouts using the red (633 nm) and the green (532 nm) laser are compared in Fig.~\ref{fig:supp:red_green}. The magnitude of the response is smaller for the green laser, but after appropriately scaling the latter, the curves in Fig.~\ref{fig:supp:red_green}(a) overlap for large detunings, where the readout is linear. The driving power of 0 dBm is higher than the highest power in Fig. 1(b) in the main text. Now, instead of a single dip, an even more complex resonance shape is visible. Instead of a dip in the center, the 1f responses vanish completely at two frequencies yet leaving a finite response in the middle. At the locations where the response goes to zero, the phase jumps by $\pi$, indicating that this corresponds to a sign change. Note that the jump around \SI{1040}{Hz} (indicated by the blue shading) corresponds to the Duffing jump from the high to the low amplitude branch (Sec.~\ref{supp:duffing}). The fact that this happens at the same detuning for the two curves indicates that the absolute motional amplitudes are identical between the two measurements; any difference between them, thus, corresponds to the readout, i.e. the laser used. Given the complex-looking phase response in Fig.~\ref{fig:supp:red_green}(b) with multiple jumps, one may ask if this still can be used to extract the mechanical nonlinearity, as we advocated in the main text. Fig.~\ref{fig:supp:red_green}(c) shows that this is indeed the case, as long as the phase data is taken modulo $\pi$. As detailed in Sec.~\ref{supp:nonlinear}, the reason for the jumps is that there are sign changes in the combined output of the nonlinear detection scheme, but the phase of the motion is unaffected. Hence by taking the data modulo $\pi$, the latter is recovered and the fitted curves match the data well. From the fitted phase of the response (Sec.~\ref{supp:duffing}), $\alpha_y/\alpha_c = 9.9 \pm 0.5$ is found for the green-laser measurement and $\alpha_y/\alpha_c = 10.0 \pm 0.6$ for the red one. From this, it is concluded that, despite the different wavelengths, laser powers, and readout efficiency, the same value for the intrinsic mechanical nonlinearity parameter can be obtained. 
Taking a closer look at the two responses in Fig.~\ref{fig:supp:red_green}(a) suggests that the green curve has a shorter ``period'' between the zeros compared to the red; the same holds for their wavelengths $\lambda$ which are  532 and 633 nm respectively. This confirms the understanding that the readout nonlinearity is caused by the motion sweeping multiple fringes; since their spacing is $\lambda/2$ the same amplitude will result in more nonlinearity in the green case, specifically a higher value of $z$ for the same amplitude $U$ (Sec.~\ref{supp:nonlinear}). This is further illustrated by the time traces in Fig.~\ref{fig:supp:red_green_time} where in the case of the green laser, there appear to be slightly more fringes just before the Duffing jump compared to the red case. % Also visible is the delay of half a period between the oscillations at large negative and large positive detuning. This can be understood using the model developed in Sec.~\ref{supp:nonlinear}.

\section{Duffing fit functions}\label{supp:duffing}
The equation of motion of the displacement of a single mode of the membrane including the cubic nonlinearity is \cite{cleland_nanomechanics}
\begin{equation}
     m_\mathrm{eff} \ddot u = -k u - m_\mathrm{eff} \gamma \dot u - k_3 u^3 + F.
    \label{eq:Duffing}
\end{equation}
Here, $k = m_\mathrm{eff}\omega_0^2$ is the linear spring constant and $k_3$ takes the Duffing nonlinearity into account. The geometric nonlinearity of rectangular membranes in discussed further in Sec.~\ref{supp:membrane_mechanics}. Furthermore, $m_\mathrm{eff}$ is the (effective) modal mass, $\gamma = 2\pi w = \omega_0/Q$ is the damping rate ($w$ is the full-width-half-maximum of the squared response $|H(f)|^2$ and $Q$ the quality factor) and $F$ is the force acting on the 
mode \cite{poot_physrep_quantum_regime}. For a harmonic driving force $F(t) = m_\mathrm{eff} \mathcal{F} \cos(\omega t)$ and by writing $u(t) = \halfl A \exp(i\omega t) + \halfl A^* \exp(-i\omega t)$ and applying the rotating wave approximation, one can express the complex motion amplitude $A$ in steady state as:
\begin{equation}
    A = \frac{\mathcal{F}/2}{\omega_0^2(1 + \frac{3}{4} \alpha |A|^2) - \omega^2 + i \omega \gamma} \label{eq:A}
\end{equation}
Here, $\alpha = k_3/k$ is the Duffing parameter with dimensions per unit length squared.
In the Lorentzian approximation (i.e. assuming that $|\omega - \omega_0| \ll \omega_0$, $Q \gg 1$, and $\alpha |A|^2 \ll 1$) and after switching from angular ($\omega$, $\omega_0$ and $\gamma$) to regular frequencies ($f$, $f_0$, $w$) one obtains:
\begin{equation}
    A \approx \frac{1}{4\pi^2}\frac{\mathcal{F}/2f_0}{f_0(1 + \frac{3}{8}\alpha |A|^2) - f + i w/2} \label{eq:AL}
\end{equation}
In the experiment, the network analyzer/lock-in amplifier does not measure the displacement $u$ (or $A$) directly, but the frequency response
\begin{equation}
    y(f) = \frac{V_\text{meas}(f)}{V_\text{out}(f)} \label{eq:y}
\end{equation}
instead. For linear transduction (nonlinear readout will be covered in Sec.~\ref{supp:nonlinear}) and excitation, there is a proportionality between the measured voltage and the motion amplitude: $V_\text{meas}(f) = \pderl{V}{u}~\xi(X,Y) A(f)$, as well as between the driving voltage and the (inertial) force $\mathcal{F} = (2\pi f)^2 p V_\text{out}(f)$ so that $y(f) \propto A(f)/\mathcal{F}$. Here, $\xi(X,Y)$ is the mode shape at the laser spot position $(X,Y)$ and $p$ takes the responsivity of the piezo and its excitation efficiency to the eigenmode into account \cite{hoch_MM_mode_mapping}. Note that the piezo-electric element may have its own response with resonances, making $p$ possibly frequency dependent. These are, however, typically much broader than the membrane resonances and, hence, $p$ is assumed to be constant. 
Combining all this leads us to use the following fit function for the driven response:
\begin{equation}
    y(f) = \frac{y_\mathrm{max}w/2}{f_0(1 + \frac{3}{8}\alpha_y |y|^2) - f + iw/2} e^{i\varphi-2\pi i (f-f_0) \tau}.
    \label{eq:Duffingy}
\end{equation}
The fit parameters are $f_0$, $w$, $y_\mathrm{max}$, $\alpha_y$, $\varphi$ and $\tau$. Note that the parameters $f_0$, $w$, $y_\mathrm{max}$ are the resonance frequency, linewidth and maximum response in the linear regime, respectively, but that these can also be determined from the nonlinear response as Eq.~\eqref{eq:Duffingy} shows. The last two parameters take into account the overall phase and a possible (group) delay, respectively. 
Note that the denominator still contains $|y|^2$ and $y$ is solved from Eq.~\eqref{eq:Duffingy} using the analytical expressions for the cubic roots. Our fitting routine can be used to either fit the magnitude $|y|$, the phase $\angle y$, or the whole complex response $y$ \cite{hoch_MM_mode_mapping}. In the case of multiple stable solutions (known as the high and low amplitude branches \cite{cleland_nanomechanics}) the fitting routine selects at every frequency the branch that best matches the data. When only the phase $\angle y$ is fitted, the magnitude $y_\mathrm{max}$ is undetermined. In this case, its value is obtained after fitting $\angle y(f)$ from the magnitude of $y$ at large detunings, where the motion is small and hence the transuction is linear.

By comparing Eq.~\eqref{eq:Duffingy} with Eq.~\eqref{eq:AL} one can connect the two representations of the Duffing parameter. Their relation is $\alpha = \alpha_y |y/A|^2 = \alpha_y (y_\mathrm{max} w f_0 / \mathcal{F})^2 = \alpha_y ([y_\mathrm{max}/Q] / [\mathcal{F}/f_0^2])^2$. Both $\alpha_y$ [Fig.~\figlinecuts(e)] and $y_\mathrm{max}$ [Fig.~\figlinecuts(b)] depend on the readout position $(X,Y)$. In contrast, $\alpha$ is a property of the mode [see Eq.~\eqref{eq:Duffing}] and it should thus be independent of the probing position, which is indeed confirmed by the measurement in Fig.~\figlinecuts(f). Note that unlike $\alpha$, $\alpha_y$ is dimensionless. As a reminder that it is obtained from the response and how it scales, we often give it the units of $\un{(V/V)^{-2}}$. Likewise, the dimensionless response $y$ and its peak value $y_\mathrm{max}$ are given in $\un{V/V}$.

\begin{figure}[tb]
  \includegraphics[width=1.0\columnwidth]{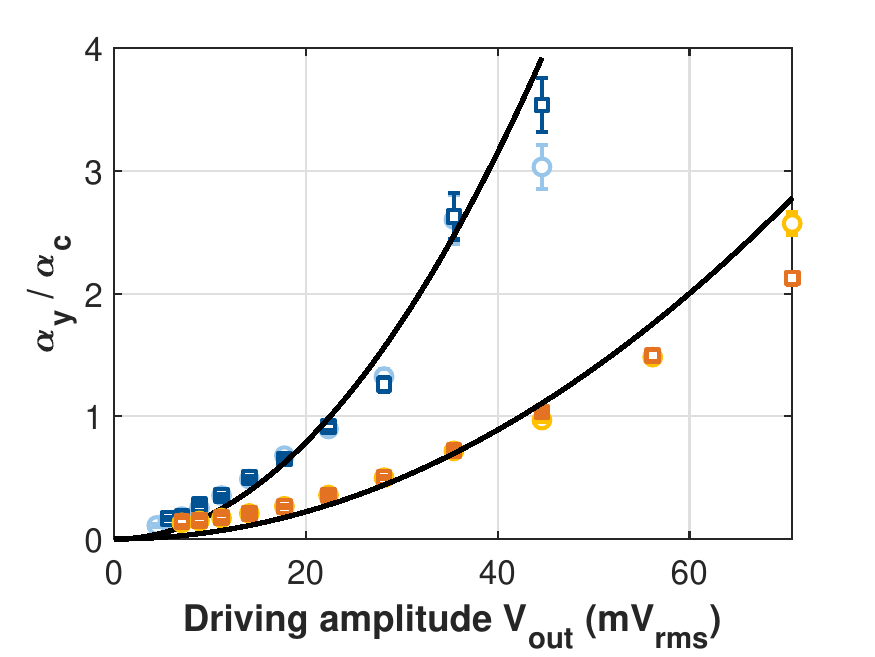}
  \caption{
  Excitation power dependence of the normalized Duffing parameter for the (1,1) mode [yellow shades] and the (2,2) mode [blue shades] at two locations on the membrane. The darker (lighter) colors are at locations with a low (high) amplitude; the colors are consistent with those from Fig.~2 of the main text. The black lines show the expected quadratic dependence from Eq.~\eqref{eq:alphanorm}.
  \label{fig:nwapower}}
\end{figure}

Depending on the value of $\alpha_y$, the solution of Eq.~\eqref{eq:Duffingy} is a Lorentzian ($\alpha_y = 0$) or develops a characteristic ``shark fin'' shape where one side of the resonance is steeper than the other ($\alpha_y \neq 0$). The critical value $\alpha_c$ is the lowest value of $|\alpha_y|$ where $|y|$ just becomes vertical \cite{cleland_nanomechanics} and in the Lorentzian approximation, its value is $ \alpha_c =  \frac{32}{27} \sqrt{3} w / (f_0 y_\mathrm{max}^2)$ \cite{nayfeh_nonlinear}. The ratio $\alpha_y/\alpha_c$ is:
\begin{eqnarray}    
    \frac{\alpha_y}{\alpha_c} & = & \frac{\alpha \left(\mathcal{F}/(y_\mathrm{max} f_0 w ) \right)^2}{\frac{32}{27} \sqrt{3} w / (f_0 y_\mathrm{max}^2)} \nonumber \\
    & = &
    \frac{\alpha \mathcal{F}^2}{\frac{32}{27} \sqrt{3} w^3 f_0} = \alpha Q^3 \frac{27}{32\sqrt{3}} \left(\frac{\mathcal{F}}{f_0^2}\right)^2.\label{eq:alphanorm}  
\end{eqnarray}
This ratio is independent of $y_\mathrm{max}(X,Y)$ and, thus, does not depend on the readout position. Its value is directly proportional to the intrinsic mechanical nonlinearity parameter $\alpha$. Figure~\ref{fig:nwapower} verifies the quadratic dependence on $\mathcal{F} \propto V_\mathrm{out}$.

\section{Multilayer reflection model}\label{supp:multilayer}
The reflection spectra from Fig.~\figtime(e) and (f) were calculated using a {\sc Matlab} program that finds the solution of the light fields $a_j^{\pm}$ inside a multilayer stack of dielectric materials. At every interface $j = 1 .. N$, Fresnel coefficients for the reflection $r_j$ and transmission $t_j$ relate the incoming fields (from the top traveling downward (-), and from the bottom traveling upward (+)) to the outgoing ones.
% Can put eqn here and/or picture from Model calculation.pdf
Between the interfaces, the fields pick up a phase $\intd \phi = \mp 2\pi n_j / \lambda \, \intd z$ while propagating a distance $\intd z$ through the material. Here, $\lambda$ is the free-space wavelength and $n_j$ is the refractive index of j-th material in the stack [see Table~\ref{table:MembraneThickness} for the values used for Fig.~\figtime(e) and (f)]. The relation between all the fields is written as a matrix equation and by fixing the incident light field, i.e. $a_0^-$ (and the one from the bottom, $a_N^+$, which is set to 0), the other fields can be solved numerically, including the upward-traveling outgoing field $a_0^+$. The latter determines the (power) reflectivity $R = |a_0^+/a_0^-|^2$. This procedure is repeated for different wavelengths $\lambda$ [Fig.~\figtime(e)] or for different distances between the SiN and the Si wafer [Fig.~\figtime(f)]. The latter is performed by adapting the distance between the Si and the SiN from the nominal value listed in Table~\ref{table:MembraneThickness}. 
For all calculations, normal incidence is assumed. 
\begin{table}
\centering
\caption{Values used in the reflectivity calculations in Fig.~\figtime. The refractive index ($n$) values used in the model include material dispersion and are from \url{refractiveindex.info}~\cite{Polyanskiy2024_refracticveindex_info}; the original work from which these values originate is also given. The last two columns give the values of $n$ at the wavelength of the red and green laser, respectively. Note that the silicon nitride is $\sim130\un{nm}$ thinner, and the inner layer of the vacuum is thicker than the stated thicknesses of the SiN and silicon oxide layers given in Sec. \ref{supp:fabrication} because of finite selectivity of BHF to silicon nitride \cite{poot_ST_AlN_simulations}.}
\label{table:MembraneThickness}
\begin{tabular}{p{20mm}|p{20mm}p{20mm}p{20mm}}
\hline 
{\bf Material}  & {\bf Thickness } &$\mathbf{n}(632.8\un{nm})$& $\mathbf{n}(532\un{nm})$ \\
\hline
\hline
Vacuum & $\infty$ (top)  & 1 & 1 \\
\hline
SiN  \cite{Philipp1973_sin_optical_prop}& \SI{183.75}{nm} &2.0395& 2.0559 \\
\hline
Vacuum & \SI{3226.25}{nm} & 1 & 1 \\
\hline
Si  \cite{Schinke2015_Si_optical_prop}& $\infty$ (bottom)  & 3.8640-0.0158i& 4.1366-0.0338i \\
\hline
\end{tabular}
\end{table}
%source
%SiN  http://refractiveindex.info/?shelf=main&book=Si3N4&page=Philipp 
% https://iopscience.iop.org/article/10.1149/1.2403440

%Si https://refractiveindex.info/?shelf=main&book=Si&page=Schinke
% https://pubs.aip.org/aip/adv/article/5/6/067168/650/Uncertainty-analysis-for-the-coefficient-of-band

\section{Nonlinear readout model} \label{supp:nonlinear}
In this section, we look at the generation of harmonics for a general nonlinear detection scheme. For this purpose, the output signal $v(t) = f\big(u(t)\big)$ is written as a (nonlinear) function of the displacement, which is assumed to be sinusoidal: $u(t) = U \cos(\omega t + \phi) = \halfl(Ae^{i\omega t} + A^*e^{-i\omega t})$. Here $U$ is the amplitude and $\phi$ is the phase of the oscillations. $A = Ue^{i\phi}$ is the complex amplitude \cite{poot_physrep_quantum_regime}. We use two models to understand how the function $f(u)$ (not to be confused with the driving frequency $f$) generates the harmonics: a power series expansion and a sinusoidal approximation. The former is especially useful for weak nonlinearities, whereas the latter is a good approximation for our interferometric optomechanical readout. In both cases, analytical expressions for the amplitudes of the harmonics are obtained.

\subsection{Power series} \label{supp:power}
A linear transduction means that the function $f(u)$ is linear in $u$. When the readout is weakly nonlinear, the function $f(u)$ can be written as a Taylor series with only a few powers in $u$ contributing \cite{dolleman_APL_calibration_NL_readout}. The more nonlinear $f(u)$ is, the more powers should be included in the expansion. 
As our first model for the nonlinear readout and the harmonics it generates, the function $f(u)$ is written as a Taylor series around $u=0$:
\begin{equation}
    f(u) = \sum_{m=0}^\infty \frac{f^{(m)}|_{u=0}}{m!} u^m.
    \label{eq:taylorf}
\end{equation}
The different powers of $u$ will generate harmonics:
\begin{equation}
   u^m = \left(\frac{U}{2}\right)^m \sum_{k=0}^m \left( \begin{array}{c}  m \\ k \end{array} \right) \exp\big(i(\omega t + \phi)(m-2k)\big). \label{eq:um}
\end{equation}
By writing $v(t)$ as a Fourier series
\begin{equation}
    v(t) = \sum_{n=-\infty}^\infty v_n e^{in\omega t} = v_0 + \sum_{n=1}^\infty 2|v_n| \cos(n\omega t + \angle v_n), \label{eq:fourierv}
\end{equation}
the Fourier coefficients are found: 
\begin{equation}
    v_n = e^{in\phi} \sum_{m=0}^\infty  f^{(m)}  \left(\frac{U}{2}\right)^m B\big(m, \halfl(m-n)\big)\label{eq:harmtaylor},
\end{equation}
where 
$$
B(m, k) = \frac{1}{m!}\left( \begin{array}{c}  m \\ k \end{array} \right)  = \frac{1}{(m-k)!k!}
$$
whenever $k = \halfl(m-n)$ is an integer between $0$ and $m$, and $B(m, k) = 0$ otherwise.
The second part of Eq.~\eqref{eq:fourierv} shows that the Fourier coefficients $v_n$ are directly related to the harmonics, in particular that the amplitude of the $n$-th harmonic is $2|v_n|$.

\begin{figure}[tb]
  \includegraphics[width=.8\columnwidth]{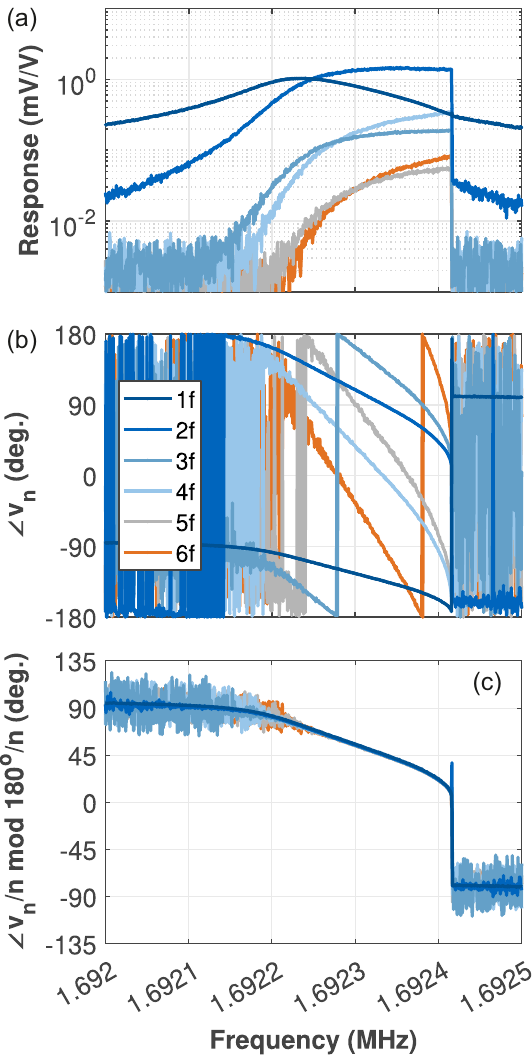}
  \caption{
  Driven response measured on the same device shown in Fig.~\ref{fig:supp:map_model_hex} with (a) magnitude and (b) phase of the first 6 harmonics. (c) shows the data from panel (b) after dividing by $n$. The higher harmonics $n\ge2$ are shifted by an integer $p$ times $180^o/n$, where $p$ is choosen to create an unwrapped curve close to that of the 1f signal.
  \label{fig:supp:phase_harmonics}}
\end{figure}

All terms except the exponential $e^{in\phi}$ in Eq.~\eqref{eq:harmtaylor} are real, which means that the phase of the $n$-th harmonic of the output signal $\angle v_n$ equals $n\phi$, possibly with $\pi$ added, depending on the sign of the outcome of the summation \footnote{Signals at different harmonics may pick up different additional phases due to the response of the readout chain of the setup including e.g. delays due cable lengths.}. This places our findings that the phase of the first harmonic modulo $\pi$ is equal to the phase of the displacement $\phi$ and that this quantity can thus be used to determine the Duffing parameter $\alpha$ on a solid theoretical footing. Moreover, the generic model of the nonlinear readout shows that the phase of all harmonics is fixed, which is an important requirement for e.g. comb generation \cite{Jong2023_mechOvertoneFreqComb}. We also observe this in the measurements. Figure~\ref{fig:supp:phase_harmonics} shows magnitude (a) and phase (b) of the first 6 harmonics as acquired during a driven response measurement with the lockin. When the amplitude rises well above the noise floor, the phase is well-defined; otherwise it displays random values between -180 and +180 deg. There is a $\sim 150 \un{Hz}$ range between 1.9925 MHz and the Duffing jump around 1.6924 MHz, where all harmonics show a well-defined value. Our model shows that $\angle v_n = n\phi~ (~+~\pi)$ and thus $\angle v_n/n$ should match $ \phi \text{~mod~} \pi/n$. Fig.~\ref{fig:supp:phase_harmonics}(c) shows that this is indeed the case: after dividing the measured phase and taking this modulo $\pi/n$, all harmonics collapse onto a single curve \footnote{We note that dividing by $n$ also reduces the variance for a randomly-distributed phase which also brings the curves closer together. This, however, is a trivial consequence. The important part is that the curves match when all phases are well defined.} that matches the phase response of the driven Duffing resonator $\phi(f)$, thus experimentally verifying the predicted fixed phase relation between the harmonic components.

Another observation is that $m-n$ has to be an even number in order to have a nonzero $B$. This means that odd powers in the Taylor expansion (i.e. odd $m$) only generate odd harmonics (i.e. odd $n$), whereas even powers generate even harmonics. Also note that 
$$
B(m, \halfl(m-n)) = \frac{1}{[\halfl(m-n)]![\halfl(m+n)]!}
$$
and that the requirement that $k = \halfl(m-n)$ is an integer between $0$ and $m$ implies that $-m \le n \le m$. The $m$-th term in the Taylor expansion thus generates harmonics $n$ up to $n=m$. The reverse question can also asked: given a harmonic $n$, which terms $m$ contribute? For this we note that, since $B = 0$ for $\halfl(m-n) < 0$, Eq.~\eqref{eq:harmtaylor} can be rewritten as:
\begin{eqnarray}    
    v_n &= &e^{in\phi} \sum_{m=n}^\infty  f^{(m)}  \left(\frac{U}{2}\right)^m B\big(m, \halfl(m-n)\big)\label{eq:harmtaylor2} \\
        &= &e^{in\phi} \sum_{p=0}^\infty  f^{(n+2p)}  \left(\frac{U}{2}\right)^{n+2p} B\big(n+2p, p\big).\label{eq:harmtaylor3}
\end{eqnarray}
Hence, it is clear that one needs a high enough power $m \ge n$ to be present in the Taylor expansion in order to reach a certain harmonic $n$. The $n$-th harmonic thus contains terms in $U$ of at least power $n$. This is also why higher harmonics have a steeper rise in Fig.~\figmaps(b). The analytical solutions of the harmonics [cf. Eq.~\eqref{eq:harmtaylor}] of the power series model thus provides insight into what powers can generate what harmonics. If the Taylor expansion of $f(u)$ converges, it is also an exact description. Still, it is noted that determining the coefficients $f^{(m)} = \partial^m f/\partial u^m|_{u=0}$ from the data turns out to be difficult as we find that the coefficients obtained from the data do not seem to converge. Looking back at Fig.~\figtime(f) of the main text shows that this is not so surprising as the reflectivity-versus-displacement fringes $R(u) \propto f(u)$ look much more like sinusoidal functions than low-order polynomials. This means that the Taylor expansion requires many terms to faithfully describe the function $f$; determining those coefficients $f^{(m)}$ from the data, e.g. using linear fitting, is typically an ill-posed numerical problem, which we will not pursue further. Instead, the second model uses a sinusoidal approximation for $f(u)$ as detailed in the following section. Interestingly, Ref.~\cite{dolleman_APL_calibration_NL_readout} took the opposite approach of using a Taylor expansion up to the 4th order to approximate a sinusoid fringe and extract the actual amplitude. In our case, due to the larger motion, this is not adequate as argued above. 

\subsection{Sinusoidal fringes} \label{supp:sine}
The periodicity of the $R(u)$ fringes in Fig.~\figtime(f) is half the wavelength $\lambda$ and they appear roughly sinusoidal. Hence, as an approximation, the output $v = f(u)$ is written as:
\begin{equation} \label{eq:fringe}
    v\big(u(t)\big) = V_0 + V_C \cos \left(2\pi \frac{u(t)}{\lambda/2}+ \theta \right)
\end{equation}
where $V_0, V_C, \theta$ describe the mean, amplitude and shift of the fringe, respectively. The linear responsivity is $\partial{v}/\partial{u}|_{u=0} = -4\pi V_C \sin(\theta) /\lambda$. By performing the Jacobi-Anger expansion, the harmonics are obtained in terms of Bessel functions of the first kind $J_n(z)$:
\begin{eqnarray*}    
     V_C \cos(\theta) J_0(z) + V_0 & \hspace{5mm} & n=0\\
    2V_C \sin(\theta) J_n(z) (-1)^\frac{n+1}{2} \cos(n(\omega t + \phi)) & & n \text{ odd} \\
    2V_C \cos(\theta) J_n(z) (-1)^\frac{n}{2}  \cos(n(\omega t + \phi)) &  & n \text{ even}
\end{eqnarray*}
where $z = 4\pi U /\lambda$. The Fourier coefficients [cf. Eq.~\eqref{eq:fourierv}] are thus 
% $v_n = V_C \sin(\theta) J_n(z) (-1)^\frac{n+1}{2} e^{i n\phi}$ for odd harmonics and likewise for even values of $n$. 
\begin{equation}
\label{eq:vnmodel}
v_n = \left\{
\begin{array}{ll}
     V_C \cos(\theta) J_0(z) + V_0 & n = 0 \\
     V_C \sin(\theta) J_n(z) (-1)^\frac{n+1}{2} e^{i n\phi} &  n \text{ odd} \\
     V_C \cos(\theta) J_n(z) (-1)^\frac{n}{2} e^{i n\phi} &  n \text{ even}
\end{array}
\right.
\end{equation}
Connecting to the results of the power-series model from Sec.~\ref{supp:power}, we note that also here $\angle v_n \mod \pi = n\phi$ and that the leading term in the asymptotic form of $J_n(z)$ for small $z$ is $\propto z^n$ and, thus $\propto U^n$.

\begin{figure}[tb]
  \includegraphics[width=1.0\columnwidth]{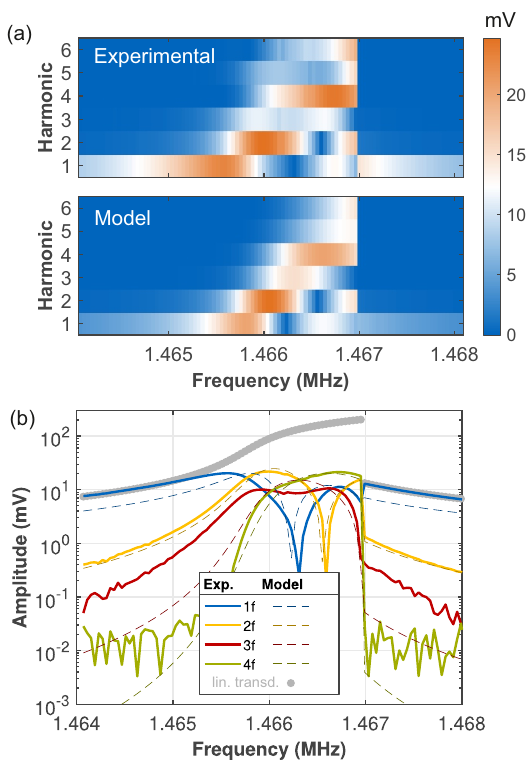}
  \caption{
  (a) Measured (top) and reconstructed (bottom) amplitude of the first 6 harmonics $2|v_n|$. The bottom panel is generated using the sinusoidal-fringe model of Eq.~\eqref{eq:fringe}. 
  % The lines show the location of cuts presented in Figure~4(e),(f). 
  (b) Line traces of the measured (solid) and reconstructed  (dashed) amplitude of the first four harmonics. The gray dots show the scaled magnitude of the phase response fit of Fig.~\ref{fig:supp:red_green}(c). These represent the 1f signal expected for a completely linear readout. The raw data underlying this Figure was measured with the red laser and was already presented in Fig.~\ref{fig:supp:red_green_time}(a).
  % nonlinear_playground_nonlinear_readout.m
  \label{fig:supp:nonlinear_nonlinear}}
\end{figure}

This simple model for the nonlinear readout describes the observed effects surprisingly well. Figures~4(c), (d) showed the amplitudes of the different harmonics in two of the recorded time traces. Such an extraction of the harmonics can be done for the whole measurement, where the excitation frequency $f$ was swept in 101 steps and time traces were recorded for every excitation frequency. The result of their decomposition is shown as the top colormap in Fig.~\ref{fig:supp:nonlinear_nonlinear}(a). The first few harmonics are also shown as line plots in  Fig.~\ref{fig:supp:nonlinear_nonlinear}(b) 
\footnote{With the bias-T in the setup (see Fig.~\ref{fig:setup}), the DC signal ($n = 0$) takes a different path compared to the high-frequency signals ($n \ge 1$) and does not arrive at the oscilloscope. Instead, the DC signal in the time traces is due to offsets and therefore the extracted $n = 0$ component is not considered}. 
For comparison, also the scaled magnitude calculated using the fit from Fig.~\ref{fig:supp:red_green}(c) is shown as gray dots in Fig.~\ref{fig:supp:nonlinear_nonlinear}(b). This represents the (1f) signal that would be generated for a completely linear readout of the nonlinear resonator. Detuned from the resonance, the first harmonic (blue) matches well with the gray dots, as the relatively low amplitude is transduced linearly. On resonance, however, the 1f signal is significantly reduced compared to the expected signal for linear readout, i.e., the blue curve lies far below the gray dots. 

With the Duffing response fitted to the phase [dotted line in Fig.~\ref{fig:supp:red_green}(c)], the model for the nonlinear readout developed in this section can be used to calculate the harmonic amplitudes as $y(f) \propto A(f)$ and, thus, $|y| \propto U \propto z$. In this case, there are three free parameters, as - in addition to $\theta$ and $V_C$ - there is also the scaling factor $S$ between $z(f)$ and $|y(f)|$. The bottom panel in Fig.~\ref{fig:supp:nonlinear_nonlinear}(a) and the dashed curves in Fig.~\ref{fig:supp:nonlinear_nonlinear}(b) show the result. Despite the simplicity of the sinusoidal-fringe model, it adeptly captures the harmonics. The overall shape of the colormaps in Fig.~\ref{fig:supp:nonlinear_nonlinear}(a) looks similar and also the lineshape in Fig.~\ref{fig:supp:nonlinear_nonlinear}(b) with its zeros is reproduced. Looking more closely at the data and model shows that the exact location of the zeros is slightly different and the lineshape of the 3f harmonic is also not reproduced perfectly. Also, the reconstructed 1f curve is offset for large detunings, indicating that the fit predicts a slightly different linear transduction factor. We believe that by extending the single-sinus model [cf. Eq.~\eqref{eq:fringe}] with a more accurate function $f(u)$, also these details of the frequency dependence of the harmonics can be captured. 

\begin{figure*}[tb]
  \includegraphics[width=1.0\textwidth]{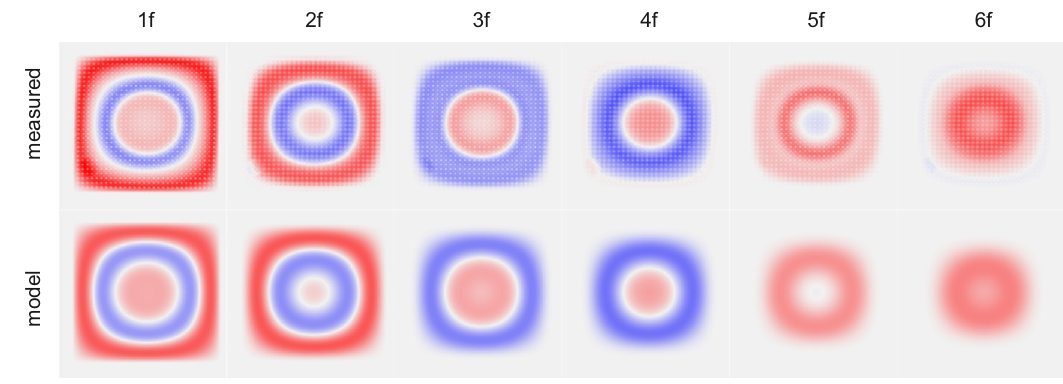}
  \caption{
  Comparison between the measured harmonics maps and the sinusoidal-fringe model as measured with the red laser. The top row shows the measured maps as already shown in Fig. 3(c) of the main text. The bottom row shows the maps that were calculated using the model. The six columns contain the 6 harmonics 1f..6f as indicated.
  \label{fig:supp:map_model}}
\end{figure*}

In addition to the frequency dependence, the model can also be used to predict the spatial profile of the different harmonics. In this case, $z$ depends on the position through the mode shape, for which we use the analytical expression of Eq.~\eqref{eq:u} and $z(X,Y) = S'\xi_{(1,1)}(X,Y)$. Again, $S'$ is another scaling factor, now between the modes shape $\xi$ and $z$, that is fitted to the experimental data in addition to $\theta$ and $V_C$. Figure \ref{fig:supp:map_model} shows the resulting maps for the first 6 harmonics; the first 3 were already shown in Fig.~\figmaps~of the main text. Again, there is a very good agreement between the measured maps and the fitted ones. The ring-like structure with its nodes (white) and regions with in-phase (red) and in anti-phase (blue) motion is captured well; only in the details like the red profiles in the 5f and 6f maps, small deviations appear. Since mode shapes are normalized such that their maximum value is 1 [Eq.~\eqref{eq:u}], the fitted value of $S' = 8.85 $ also corresponds to the maximum value of the modulation parameter $z$, which occurs at the center of the membrane where the anti-node of the (1,1) mode is located.

\begin{figure*}[tb]
  \includegraphics[width=1.0\textwidth]{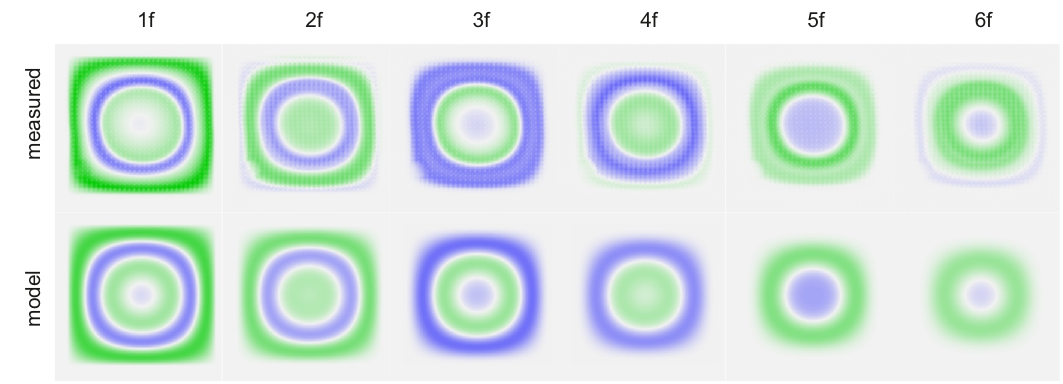}
  \caption{
  Comparison between the measured harmonics maps and the sinusoidal-fringe model as measured with the green laser. The top row shows the measured maps, whereas the bottom row shows the maps that were calculated using the model. The six columns contain the 6 harmonics 1f..6f as indicated.
  \label{fig:supp:map_model_green}}
\end{figure*}

The same measurement has also been done with the green laser (see Fig.~\ref{fig:supp:map_model_green}). In this case, the maximum value of $z = 10.51$ is higher compared to the measurement with the red laser where the maximum was $z = 8.85$. Their ratio is $1.187$, which is in very good agreement with the ratio of the wavelengths $632.8\un{nm}/532\un{nm} = 1.190$ as predicted by the definition of $z$, further supporting our model for the readout nonlinearity. Note that the values for $z$ correspond to calibrated \cite{dolleman_APL_calibration_NL_readout} center amplitudes of $U = 446 \un{nm}$ and $445 \un{nm}$ for red and green, respectively, which are in good agreement with each other.
% unrounded: 445.65, 444.9 nm

%

\section{Membrane mechanics}
\label{supp:membrane_mechanics}
The eigenfrequencies $f_{(m,n)}$ of the flexural modes of a rectangular membrane of size $L_x$ by $L_y$ under large, uniform, and isotropic tension are: 
\begin{equation}
    f_{(m,n)} = \frac{c}{2}\left( {\left(\frac{m}{L_x}\right)^2 + \left(\frac{n}{L_y}\right)^2 }\right)^{1/2}. \label{eq:f}
\end{equation}
The associated mode shapes are:
\begin{equation}
    \xi_{(m,n)}(X,Y)  = \sin(\pi m X / L_x) \sin(\pi n Y / L_y). \label{eq:u}
\end{equation}
Here, $m$ and $n$ are the number of antinodes in the $x$ and $y$ direction, respectively \cite{hoch_MM_mode_mapping}. For flexural vibrations, the speed of sound is $c = (\sigma/\rho)^{1/2}$, where $\rho$ is the mass density and $\sigma$ is the stress. The dimensions of the studied membrane are $L_x = L_y = 270 \un{\mu m}$ (Sec.~\ref{supp:fabrication}).

We note that fully capturing all mechanics details of our membranes, including periodically-arranged release holes and etch effects \cite{adiga_APL_SiN_drum_Q_mode, sommer_APL_membrane_AlN} requires numerical simulations, in particular mechanical band structure calculations as detailed in Refs.~\cite{sommer_APL_membrane_AlN, poot_ST_AlN_simulations}. From such an analysis, the stress in the silicon nitride was found to be $\sigma = 1101.5 \un{MPa}$ and $c = 559.0 \un{m/s}$. Another conclusion was that effects like bending rigidity, release holes, finite selectivity, etc. are only small corrections and, hence, to a very good approximation, the devices behave as ideal membranes under tension. 

\subsection{Geometric nonlinearity of a rectangular membrane}
Vibrational eigenmodes of a rectangular membrane are a classic problem in mechanics and also their nonlinearities have been explored. Both positive and negative displacements of the membrane lead to increased stretching beyond the initial strain. On average, the tension increases, and with it also the resonance frequency. This so-called geometric nonlinearity is the cause of the Duffing response observed in the membranes. The parameter $\alpha$ is thus determined by the membrane geometry and by the specific mode that is studied.
For e.g., a one-dimensional string this geometric effect is easily solved \cite{yao_PRB_predisplaced}, but the two-dimensional case considered here is more involved because the displacement-induced stress is not uniform~\cite{Cattiaux_JAP_geometric_NL_drum}. Ref.~\cite{liu_IAM_geometric_membrane} gives an overview of the existing literature and gives the derivation for a rectangular membrane. Their final result (for an isotropic membrane and in our notation) is:
\begin{equation}
    \frac{k_3}{m_\mathrm{eff}} = \frac{3\pi^4}{16} \frac{E}{\rho}\left(\frac{m^4}{L_x^4}+\frac{n^4}{L_y^4}\right).
\end{equation}
Together with 
\begin{equation}
    \frac{k}{m_\mathrm{eff}} = \omega_{(m,n)}^2 = \big(2\pi f_{(m,n)} \big)^2 = \pi^2 \frac{\sigma}{\rho} \left(\frac{m^2}{L_x^2}+\frac{n^2}{L_y^2}\right)
\end{equation}
the Duffing parameter is found:
\begin{equation}
    \alpha_{(m,n)} = \frac{k_3}{k} = 
    \frac{3\pi^2}{16} \frac{E}{\sigma}\left(\frac{m^4}{L_x^4}+\frac{n^4}{L_y^4}\right)/\left(\frac{m^2}{L_x^2}+\frac{n^2}{L_y^2}\right).
\end{equation}
For the (1,1) mode of a square membrane with $L_x = L_y = L$ one obtains $\alpha_{(1,1)} = \frac{3\pi^2}{16} \frac{E}{\sigma} \frac{1}{L^2}$ and $\alpha_{(2,2)} = 4\alpha_{(1,1)}$. Both are proportional to $L^{-2}$ and, thus, the same displacement amplitude leads to larger relative frequency shift for smaller membranes. For $L = 270 \un{\mu m}$, $\sigma = 1101.5 \un{MPa}$, $E = 250 \un{GPa}$ one obtains $\alpha_{(1,1)} = 5.7\times 10^{-3} \un{\mu m^{-2}}$. 

\section{Parametric effects}
\label{supp:parametric}
\begin{figure}[tbh]
  \includegraphics[width=0.7\columnwidth]{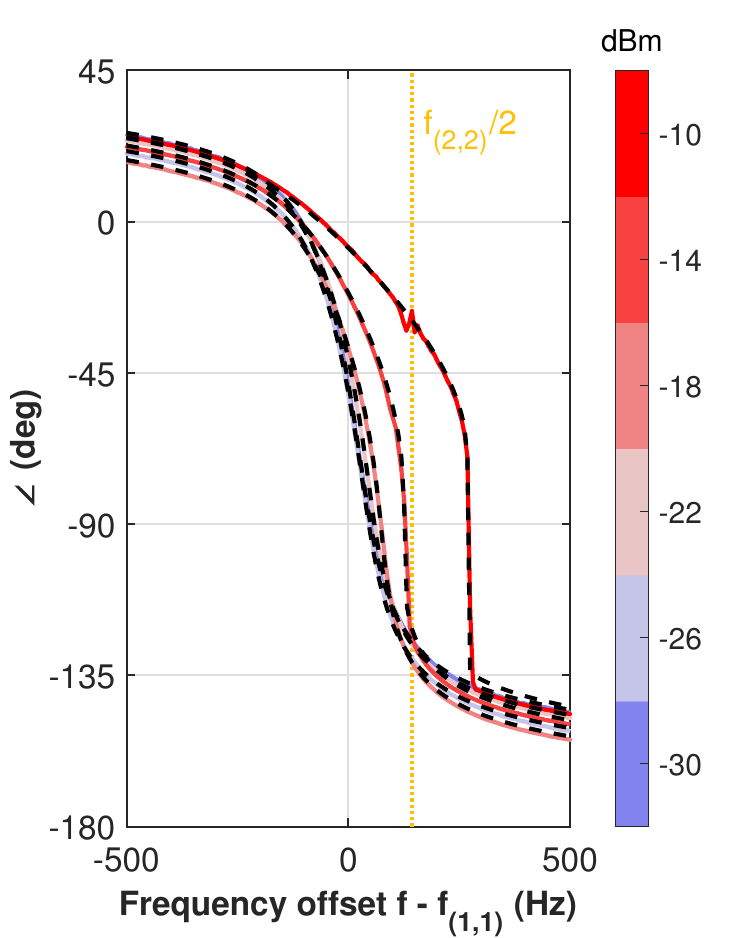}
  \caption{
  The phase of the driven (1,1) responses that were shown in Fig.~\figintro(b). The black dashed lines indicate the fitted Duffing response and the gold dotted line indicates half the frequency of the (2,2) mode. At the highest driving power, a small wiggle is visible in the measured phase response.
  \label{fig:2f}}
\end{figure}

Equation \eqref{eq:f} predicts that the frequency ratio between the two modes under study is $f_{(2,2)}/f_{(1,1)} = 2$ and parametric ``2f'' effects, such as internal resonances, (anti)squeezing, or even parametric oscillations \cite{rugar_PRL_squeezing, poot_NJP_Yfeedback, mahboob_natnano_bit, ruzziconi_NLD_internal, xiong_NLD_internal} may be possible.
Experimentally, $f_{(2,2)} = 2.932\,369 \un{MHz}$, and  $f_{(1,1)} = 1.466\,060 \un{MHz}$ \footnote{These values shifted slightly during the measurement campaign (see e.g. Fig.~\ref{fig:mode_maps}), possibly due to changes in the lab environment.}, so their ratio is not exactly 2, but $2.0017$. The small difference $f_{(2,2)} - 2 f_{(1,1)} \sim +250 \un{Hz}$ is attributed to the bending rigidity \cite{yu_PRL_membrane_Al,yang_PRL_persistent_response,poot_ST_AlN_simulations}. Note that $f_{(2,2)} - 2 f_{(1,1)} \sim 0.9~w_{(1,1)}$ is thus at least comparable (or even much larger: $\sim 30.6~w_{(2,2)}$) than the line width $w_i$ so that 2f effects are suppressed even when driving the (2,2) mode strongly. No obvious anomalies were detected in the nonlinear responses of the (2,2) mode. 
However, when taking a close look at Figure~\ref{fig:2f}, which shows the phase of the data whose magnitude was already shown in Fig.~\figintro(b), a small wiggle is visible (it is also visible - although even less pronounced - in Fig. ~\ref{fig:supp:red_green}). Its position coincides with half $f_{(2,2)}$ (gold line). Still, this wiggle does not significantly affect the fitted curves and future research will look further into this feature that is probably related to parametric effects and mode coupling as observed e.g. in Ref.~\cite{yang_PRL_persistent_response}. This will include spectrum analyzer measurements around the 2f harmonic and pump-probe measurements \cite{westra_PRL_coupled}. 

\begin{figure*}[!b]
  \includegraphics[width=0.95\textwidth]{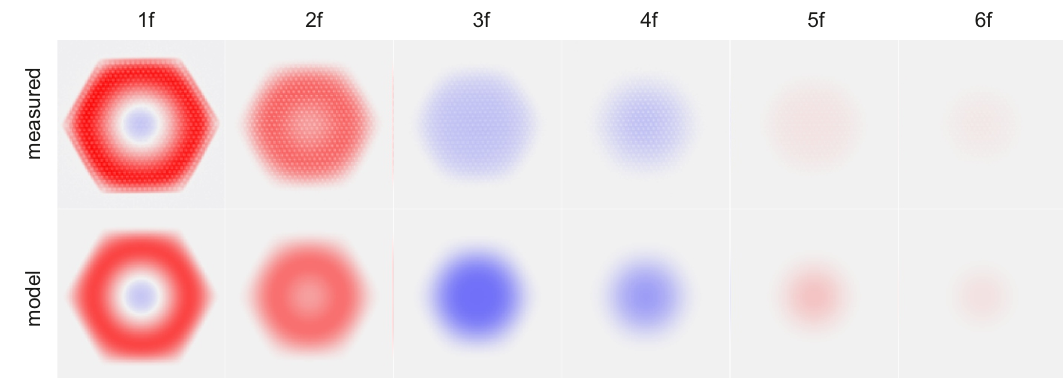}
  \caption{
  Comparison between the measured (with the red laser) harmonics maps and the sinusoidal-fringe model of the fundamental mode of a hexagonal membrane. The long axis of the hexagon measures $278.7 \un{\mu m}$. The top row shows the measured maps, whereas the bottom row shows the reconstructed maps. The six columns contain the 6 harmonics 1f..6f as indicated.
  \label{fig:supp:map_model_hex}}
\end{figure*}

Still, to fully exclude that our spatial patterns are caused by coupling to higher modes via internal resonances, we have performed measurements on a \emph{hexagonal} membrane on an independently fabricated chip. Importantly, for the hexagonal geometry the higher modes are not close to the harmonics of the fundamental resonance, thus avoiding internal resonances. Specifically, finite element simulations predict a ratio of $f_i/f_1 = 1.0000$, 1.5930, 1.5930, 2.1335, 2.1336, 2.2946, 2.5877, 2.7216, 2.9117, and 2.9118 for $i=1..10$, which we find to be in good agreement with the experimentally observed eigenfrequencies for the first 6 modes.
The fundamental mode of the hexagonal membrane is $f_1 = 1.6858 \un{MHz}$ for this $278\un{\mu m}$-sized membrane (length of the long axis) and it can be driven to the point where nonlinearities in the readout become important. 
Figure~\ref{fig:supp:map_model_hex} shows the measured harmonics maps of its strongly-driven fundamental mode. 
Simulations and maps in the small-amplitude regime show that the fundamental mode of a hexagonal membrane also does not have a node. Still, also in Fig.~\ref{fig:supp:map_model_hex} measurement, there is a clear anti-phase region (blue) in the center of the 1f map and also the other harmonics show up. There is again a good agreement between the experiments and the model (bottom row). In this case, instead of using an analytical expression, the simulated mode shape $\xi_1(X,Y)$ is used to fit the maps. The fit of the maximum modulation parameter yields $S' = 4.218$ and $U = 212 \un{nm}$. This is lower than for the square membrane studied earlier, where $S'$ was $8.85$ and $U=446 \un{nm}$. The smaller modulation of the hexagonal membrane is also manifested by the decrease in signal with increasing harmonic number as expected from Sec.~\ref{supp:nonlinear}.
Hence, the observed effects are thus independent of the sample, the specific membrane, its geometry, and mode spectrum.

\ifx\calledSI\undefined
    \bibliography{membranes}
    \end{document}
\fi

\fi

\end{document}